\documentclass{svproc}

\usepackage{url}
\usepackage{graphicx}%
\usepackage{multicol}%
\usepackage{footmisc}%
\usepackage{amsmath,amssymb,amsfonts}%
\usepackage{mathrsfs}%
\usepackage[title]{appendix}%
\usepackage{xcolor}%
\usepackage{textcomp}%
\usepackage{manyfoot}%
\usepackage{booktabs}%
\usepackage{algorithm}%
\usepackage{algorithmicx}%
\usepackage{algpseudocode}%
\usepackage{listings}%
\usepackage{geometry} 
\begin{document}







\mainmatter              
\title{Machine Learning Detection of Correlations in Snapshots of Ultracold Atoms in Optical Lattices}

\author{Stephanie Striegel\inst{1} \and Eduardo Ibarra-Garc\'ia-Padilla \inst{1,2} \and Ehsan Khatami\inst{1}}

\institute{San Jos\'e State University, San Jos\'e CA 95192, USA,\\
\email{stephanie.striegel@sjsu.edu}
\email{ehsan.khatami@sjsu.edu}
\and
University of California, Davis, Davis CA 95616, USA, \\
\email{edibarra@ucdavis.edu}}

\maketitle

\abstract{

Recent proposals have suggested the use of supervised learning with convolutional neural networks to shed light on some of the less well known phases of the Fermi-Hubbard model through the classification of snapshots from the quantum gas microscopy of ultracold atoms in optical lattices. However, there have been challenges in the interpretability of networks with more than one convolutional filter coupled to the input images. Here, we expand on previous work by considering multiple filters in the first convolutional layer and developing a process for analyzing the physical relevance of patterns obtained in the trained filters. 
We benchmark our approach at half-filling, where strong antiferromagnetic correlations are known to be present, and we find that upon hole doping, previously unknown patterns arise at temperatures below the tunneling amplitude. These patterns may be a signature of interesting arrangements of fermions in the lattice.}

\section{Introduction}\label{sec1}

The paradigmatic Fermi-Hubbard model (FHM) is one of the most studied models in condensed matter physics because of its close connection to the physics of superconducting cuprates~\cite{Arovasreview,Qinreview}. The FHM Hamiltonian is given by, 
\begin{equation}\label{eq:Hubbard_N1}
H = -t \sum_{\langle i,j \rangle, \sigma} \left( c_{i \sigma}^\dagger c_{j \sigma}^{\phantom{\dagger}} 
+ \mathrm{h.c.} \right) + U \sum_{i} n_{i \uparrow} n_{i \downarrow} - \mu \sum_{i,\sigma} n_{i \sigma},
\end{equation} 
where $c_{i \sigma}^\dagger$ ($c_{i \sigma}^{\phantom{\dagger}} $) is the creation (annihilation) operator for a fermion with spin $\sigma = \uparrow,\downarrow$ on site $i = 1,2,...,N$. $N$ denotes the number of lattice sites, $n_{i \sigma} = c_{i \sigma}^\dagger c_{i \sigma}^{\phantom{\dagger}}$ is the number operator for spin $\sigma$, $t=1$ (setting the unit of energy throughout the paper) is the nearest-neighbor tunneling amplitude, $U$ is the on-site interaction strength, and $\mu$ is the chemical potential.\\ 

Experiments using cold atoms in optical lattices are well described by Eq.~\eqref{eq:Hubbard_N1} and have become an invaluable tool to probe the FHM's physics due to their high degree of control over all models parameters \cite{a_bohrdt_21}. Furthermore, the development of quantum gas microscopy (QGM)~\cite{Bakr2009,Gross2021} for two-dimensional optical lattices has provided the possibility to directly detect long-range correlation functions through real-space and spin-resolved imaging of fermionic atoms, and have produced significant results in understanding the FHM's phase diagram~\cite{a_bohrdt_21}. However, despite their enormous success, these experiments still suffer from the inability to reach temperature regions relevant to some of the most sought after phases in the model, such as those with significant charge density wave or pairing correlations, as well as measurements that would lend themselves to detecting phases with off-diagonal or exotic order, such as the pseudogap or strange metal phases.\\

\begin{figure}
\centering
\includegraphics[scale=0.7]{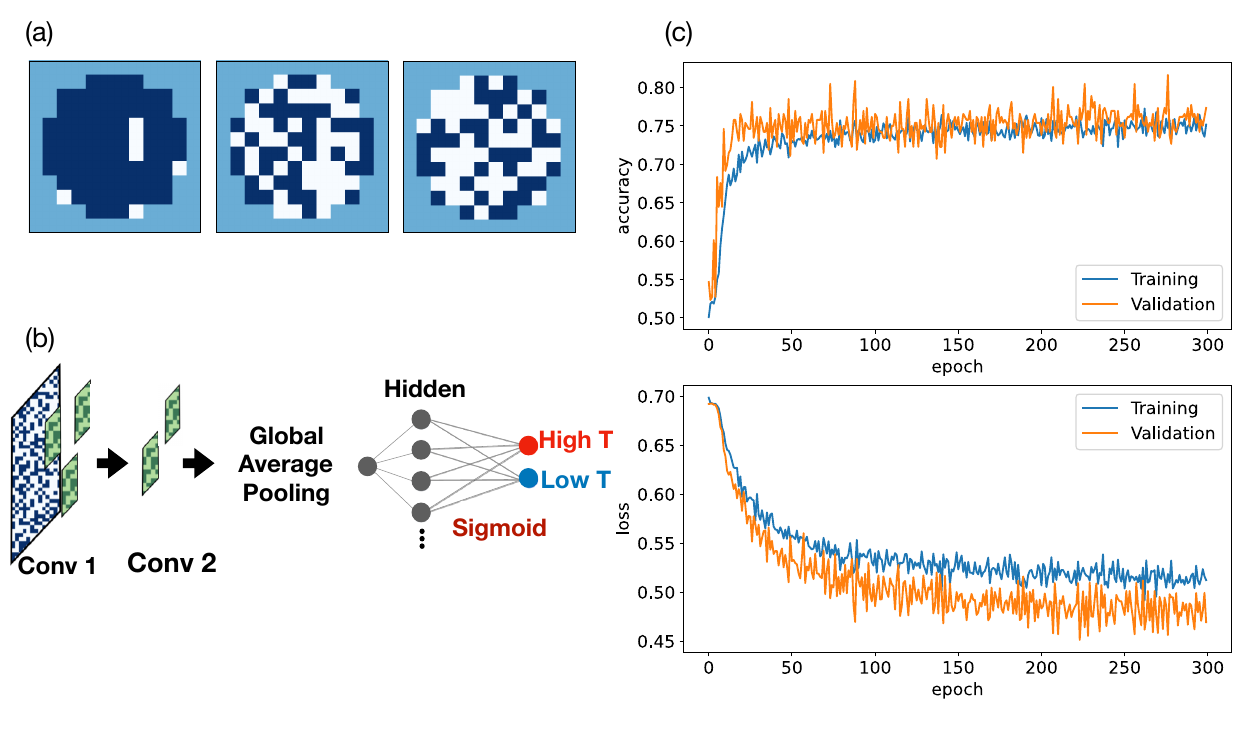}
\caption{(a) Visualization of a sample experimental data at half filling. Left: Snapshot of singles, where a dark blue pixel denotes an atom and a white pixel denotes either a hole or a double occupancy (doublon). Middle: Snapshot of spin-up atoms. Right: Snapshot of spin-down atoms. (b) Schematics of our CNN. There are two convolutional layers with three and two filters, followed by a global average pooling layer and a fully-connected layer with six neurons. The two output neurons allow for the categorization of input snapshots as belonging to high or low temperature regions. (c) Sample evolution of the accuracy and loss functions during training using snapshots at half filling.}
\label{fig:data}
\end{figure}

Machine learning tools have been used in recent studies to shed light on some of the less well known phases of the model at low temperatures through the analysis of QGM snapshots (projective measurements of density)~\cite{a_bohrdt_18,e_khatami_20,c_miles_23}.   
In this paper, we extend the work done in Ref.~\cite{e_khatami_20} in which trained filters of a simple convolutional neural network (CNN) were analyzed to infer ordering patterns of fermions in projective measurements. We use a publicly available experimental data set for spin-resolved density snapshots of the model with $U=8.1$ on the square lattice~\cite{c_chiu_19} [see Fig.~\ref{fig:data}(a)] to train a modified CNN that has multiple filters in its first convolutional layer.
We focus our attention to projective measurements at half-filling (zero doping, average density of one atom per site), and 8\% hole doping. Even though the non-linearities in the neural network architecture make the interpretation of the trained filters nontrivial, here, we ask whether it would be possible to isolate individual filters and identify the temperature region in which patterns formed in them are favored, thereby learning from them about relevant correlations in the system at low temperatures.

\section{Method}\label{subsec2}

We use a CNN to analyze the snapshots of ultra-cold atoms in an optical lattice. Although the publicly available data
we use are taken across a wide range of temperatures, we are mostly interested in the network's ability to differentiate between snapshots at `high' temperatures and those at `low' temperatures. For that reason, we only use snapshots taken at the extreme temperatures during our training. If a CNN can be trained to make the distinction reasonably well, we hypothesize that the filters in the first convolutional layer, which are directly connected to the physical snapshots, capture important patterns in the snapshots that are relevant either at low temperatures or at high temperatures.\\

Our CNN is comprised of two convolutional layers. The first layer has three $5\times 5$ filters and the second layer has two $5\times 5$ filters. Dropout layers follow the convolutional layers which are then connected to a global average pooling layer. It is then flattened and batch normalized before being fed to a hidden layer with six neurons. The output layer has two neurons with the sigmoid activation function [Fig.~\ref{fig:data}(b)]. This network is trained on 1000 spin-up and spin-down snapshots, 500 at the lowest temperatures ($T\le 0.6$) and 500 at the highest temperatures ($T\ge 1.2$). Because of the relatively small number of snapshot available for training, we implement data augmentation by applying point-group symmetries to each snapshot to increase the number of samples sixfold (using three consecutive $90^\circ$  rotations and two reflections about the horizontal and vertical axes). This improves the network accuracy by $\sim5\%$. We split the data into training and validation sets, 90\% and 10\%, respectively. The batch sizes for the training and validation sets are 32 and 8, respectively. See Fig.~\ref{fig:data} (c) for a sample evolution of accuracy and loss functions during a training involving snapshots at half filling.\\

After a training, we isolate each of the filters in the first convolutional layer and perform our own convolutions of them this time with snapshots at all temperatures using the same stride as used in the CNN. We further apply a rectified linear unit (ReLU) to the resulting convolutions and study their average as a function of temperature.

\begin{figure}[t]
\centering
\includegraphics[scale=0.7]{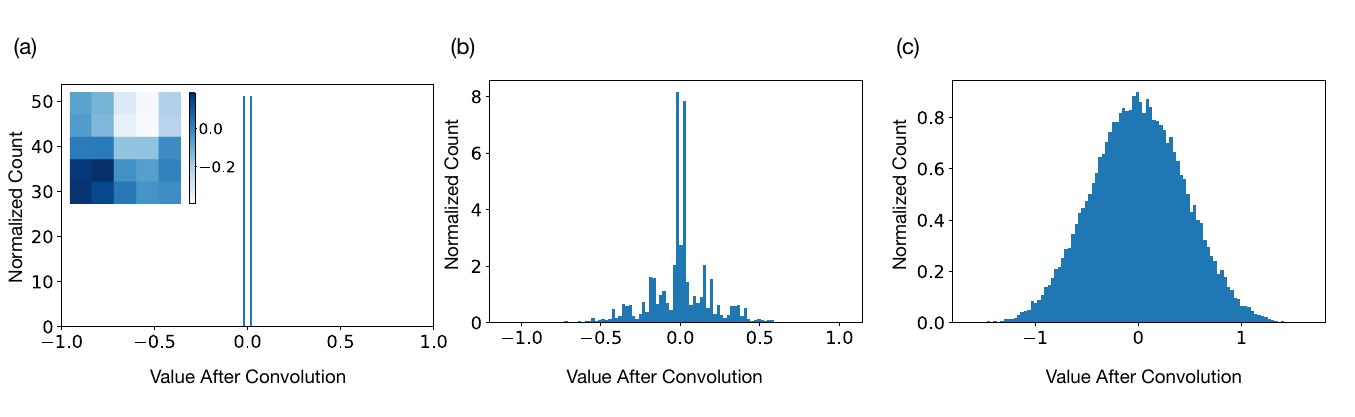}
\caption{Histograms of the average values of convolutions of a sample trained filter [shown in the inset of (a)] and (a) the `perfect' checkerboard snapshot, (b) snapshots generated after 5\% of pixels in the perfect snapshot are randomly flipped, and (c) snapshots generated after 90\% pixels in the perfect snapshot are randomly flipped. As the disorder increases, we see a Gaussian-like distribution begin to form centered around 0.}
\label{fig:hist1}
\end{figure}

\section{Results}\label{sec4}

In the CNN, a bias value is added to the results of the convolution with each filter before passing those results through the ReLU. In our manual convolutions, it is not clear a priori what we should choose as an appropriate bias value for each filter. To gain some insight, we analyze the distribution of values that result from the convolution of each filter with parts of sample snapshots with the goal of using a clearly defined mean value as our bias. To perform an unbiased analysis, we start with a synthetically generated snapshot that represents the perfect classical antiferromagnetic order at half filling (the checkerboard pattern). Then, to mimic noise and quantum fluctuations in real snapshots, we gradually introduce disorder, through flipping pixel values at random locations, and generate many samples for each disorder strength, defined as the number of flips applied to the perfect antiferromagnetic order.\\

We find that regardless of the filter, the distribution of convolutions quickly evolves from a bimodal one (expected for the perfect checkerboard structure) to one resembling a Gaussian centered around zero upon introducing disorder. Figure~\ref{fig:hist1} displays the resulting histograms for a sample trained filter shown in the inset of Fig.~\ref{fig:hist1}(a). Therefore, we conclude that a bias of 0 is probably the most appropriate one for our analysis. We note that a similar distribution as in Fig.~\ref{fig:hist1}(c) emerges when using real snapshots from the experiment.\\


\begin{figure}
\centering
\includegraphics[scale=0.9]{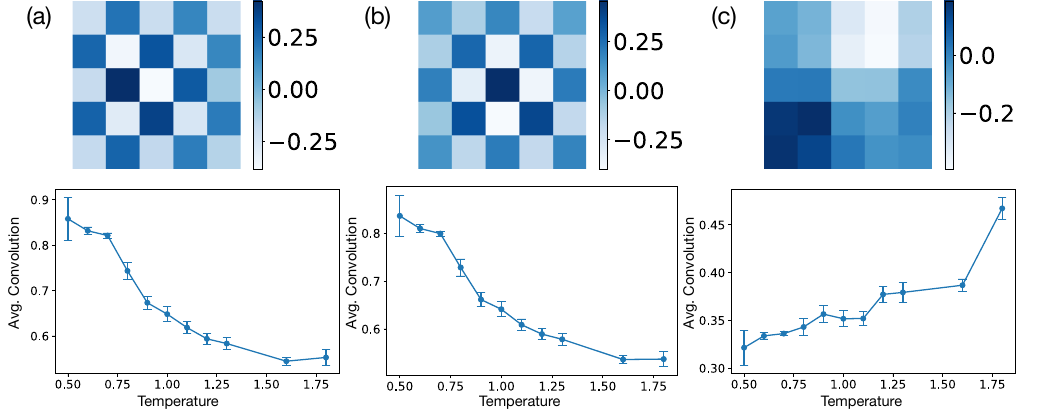}
\caption{Plots of average convolutions resulting from the manual convolution of each filter and the data at half-filling vs temperature. The three filters are from the first convolutional layer in our CNN after a typical training. Higher averages suggest that the pattern is more likely to be observed at the corresponding temperature.}
\label{fig:results_hf}
\end{figure}

\begin{figure*}
\centering
\includegraphics[scale=0.9]{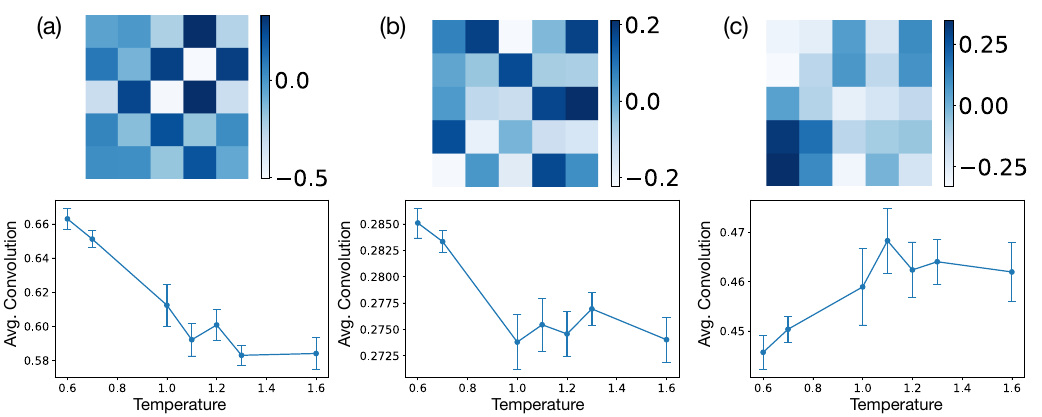}
\caption{Same as Fig.~\ref{fig:results_hf}, except that the training has been done using snapshots at 8\% doping.} 
\label{fig:results_8per}
\end{figure*}

Our main findings are summarized in Figs.~\ref{fig:results_hf} and~\ref{fig:results_8per}, where we present (1) typical filters obtained from a training of the CNN using spin-up and spin-down snapshots at half-filling (Fig.~\ref{fig:results_hf}) and at 8\% doping (Fig.~\ref{fig:results_8per}), and (2) the averages obtained after manually convolving each filter with the data sets as a function of temperature. We note that the spin does not have a preferred direction in the experimental setup, and so, we treat spin-up and spin-down snapshots as the same.\\

In the case of half filling, we find accuracies that are around $75\%$, and at least one filter in each training that exhibits patterns reflecting the antiferromagnetic order, expected to develop at half-filling at low temperatures [see Fig.~\ref{fig:results_hf}(a)-(b)]. The lower panels in Fig.~\ref{fig:results_hf}(a)-(b) show that the average convolutions for these filters are significantly larger at low temperatures than at high temperatures, clearly indicating that such order is favored in the low-temperature region. On the other hand, the third filter shown in Fig.~\ref{fig:results_hf}(c), results in convolutions that are larger at high temperatures, indicating that the patchy ferromagnetic pattern show up mostly in the high-temperature snapshots. \\

At 8\% doping, the accuracy drops to $\sim60\%$. This  could signal the presence of more complex magnetic structures away from half-filling. Filters from a typical training at this doping are illustrated in Fig.~\ref{fig:results_8per}. A shorter-range checkerboard pattern, in comparison to those seen in Fig.~\ref{fig:results_hf}, appears in the first filter shown in Fig.~\ref{fig:results_8per}(a), and is favored again at low temperatures. It points to the existence of remnant antiferromagnetic correlations that may extend to 1-2 sites at this doping, consistent with theory~\cite{Tranquada}.
The other pattern observed in the second filter in Fig.~\ref{fig:results_8per}(b), also favored at low temperatures, is more difficult to interpret physically, but hints at possible diagonal lineup of spins. Finally, the third filter shown in Fig.~\ref{fig:results_8per}(c) is less interesting as it is slightly more favored at higher temperatures. \\

\section{Conclusion}

By analyzing filters of CNNs trained to distinguish snapshots of ultra-cold atoms in optical lattices at low and high temperatures, we can find patterns that point to physical correlations favored at low temperatures in the Hubbard model. In this study, we demonstrated that training with single-spin-species snapshots at half filling, the filters clearly identify antiferromagnetism as the dominant low-temperature feature. Similar trainings with snapshots at 8\% hole doping result in firs that show antiferromagnetic correlations are shorter range than at half filling and suggest other interesting patterns favored at low temperatures. Further analysis using more sophisticated artificial neural networks that take both spin and charge snapshots for the same experimental sample as input can lead to discoveries of intertwined spin and charge orders away from half filling. 

\section*{Acknowledgments}
This material is based upon work supported by the U.S. Department of Energy, Office of Science, Office of Basic Energy Science's Data Science to Advance Chemical and Materials Sciences program under Award Number DE-SC-0022311.

\bibliography{References.bib}



\end{document}